\documentstyle[twocolumn,aps]{revtex}
\begin{document}
\draft
\title{Influence of the confinement geometry on surface superconductivity} 
\author
{V.A. Schweigert \cite {*:gnu} and F.M. Peeters \cite {f:gnu}}
\address{\it  Departement Natuurkunde, Universiteit Antwerpen (UIA),\\
Universiteitsplein 1, B-2610 Antwerpen, Belgium}
\date{\today}
\maketitle

\begin{abstract}
The nucleation field for surface superconductivity,
$H_{c3}$, depends on the geometrical shape of the mesoscopic superconducting
sample and is substantially enhanced with decreasing sample size.
As an example we studied circular, square, triangular and wedge shaped
disks. For the wedge the nucleation field diverges as
$H_{c3}/H_{c2}=\sqrt{3}/\alpha$ with decreasing angle ($\alpha$) of the wedge,
where $H_{c2}$ is the bulk upper critical field.
\end{abstract}

Recent progress in microfabrication techniques has made it possible to study
mesoscopic superconducting samples of micronmeter and submicronmeter 
dimensions\cite{mos1,geim}. The size and shape of such samples 
strongly influences the superconducting properties. 
Whereas bulk superconductivity exists at 
low magnetic field (either the Meissner state for $H<H_{c1}$ in type-I and 
type-II
superconductors or the Abrikosov vortex state for $H_{c1}<H<H_{c2}$
in type-II superconductors), surface superconductivity survives
in infinitely large bounded samples up to the third critical
field $H_{c3}\approx 1.695H_{c2}$\cite{saint}.

For mesoscopic samples of different shape,
whose characteristic sizes are comparable to the coherence
length $\xi$, recent experiments \cite{mos1,geim,buisson,strunk1,zhang}
have demonstrated the Little-Parks-type \cite{little}
oscillatory behavior of the phase boundary between the normal
and superconducting state in the $H-T$ plane, where $T$ and $H$ are the
critical temperature and the applied magnetic field, respectively.
While for a circular sample, when the superconductor state is
characterized by a definite angular momentum, the ocsillatory behavior of the
nucleation field is well understood theoretically \cite{mos2,benoist,deo,me},
the problem of the nucleation of superconductivity in arbitrary shaped systems
is still to be solved. 

Earlier numerical calculations by the present authors\cite{me}
showed that the oscillations of the
nucleation field with sample size for ellipsoidal and rectangular shaped
samples disappears only for large aspect ratios. 
Recently, the asymptotic behavior of
the nucleation field in the small size limit, $L\gg \xi$, and in the large size
limit, $L\ll \xi$, where
$L$ is the characteristic sample size, was found
for smooth (i.e. without sharp corners) and square samples \cite{jadallah}.
The particular case of surface superconductivity in a rectangular loop
\cite{fomin1} and in a wedge \cite{fomin} was investigated recently. 
In Ref.~\cite{fomin} a variational
approach was used and they found that $H_{c3}$ is maximum for a wedge angle of $\alpha
=0.44\pi$. This is surprising in view of the intuitive idea that surface
superconductivity increases with decreasing size of the system and therefore,
one would expect that $H_{c3}$ is an uniform increasing function with
decreasing $\alpha$. The result of Ref.~\cite{fomin} was also in disagreement
with calculations \cite{me,jadallah} for the particular case of
a square sample.

In the present paper, we calculate, within the Ginzburg-Landau (GL) mean field
approach, the
nucleation field for bounded square and triangular samples as well as
for a wedge with infinitely long sides and arbitrary corner angle $\alpha$.
We resolve the discrepancy between the results of Refs.~\cite{fomin} and
\cite{jadallah}. Furthermore, we obtain the following analytical expression for 
the asymptotic behavior of the nucleation field of a wedge 
$H_{c3}/H_{c2}=\sqrt{3}/\alpha$ in the limiting case of $\alpha\ll 1$.

{\it Nucleation of superconductivity in a finite bounded sample.}
Close to the superconducting - normal transition the demagnetization effects
are not important and consequently the nucleation of 
superconductivity is described by the first 
GL equation for the order parameter $\Psi$ 
\begin{equation}
\label{ne1}
\frac{1}{2m^*}\left(-i\hbar\vec \nabla-\frac{e^*}{c}\vec A\right)^2\Psi
=-\alpha\Psi-\beta\Psi|\Psi|^2,
\end{equation}
where $\alpha$, $\beta$ are the GL parameters which depend on 
temperature, $\vec A$ is the vector potential of the applied magnetic field
$\vec H=rot \vec A$ and $m^*=2m$ and $e^*=2e$ are the mass and electrical
charge of the Cooper pairs, respectively.
We consider a uniform external
magnetic field applied along the $z-$axis and assume that 
the order parameter is uniform in this direction.
The latter is satisfied for samples which are close to the nucleation field 
and which have flat top and bottom sides.
When calculating the nucleation field we neglect the nonlinear term 
and rewrite Eq.~({\ref{ne1}}) as
\begin{equation}
\label{ne2}
\left(-i\vec \nabla-\vec A\right)^2\Psi=-\Psi,
\end{equation}
where the distance is measured in units of the coherence length
$\xi=\hbar/\sqrt{-2m\alpha}$, the vector potential in
$c\hbar/2e\xi$, and the magnetic field in $H_{c2}=c\hbar/2e\xi^2$,
the bulk upper critical field.

Thus the problem of nucleation of superconductivity is reduced to the 
calculation of the lowest eigenvalue and corresponding 
eigenfunction of the two-dimensional (2D) linear operator
$\hat L=(-i\vec \nabla-\vec A)^2$ which also describes 
the motion of a quantum particle. 
The transition from the normal to the superconducting
state occurs when the lowest eigenvalue of $\hat L$ becomes smaller
than unity. Contrary to the case of the usual Schr\"odinger equation, where the
wave function is taken equal to zero at the sample boundaries, here the
normal component of the superconducting current must be zero at the
boundary between the superconducting and the insulating material
\begin{equation}
\label{ne3}
\left(-i\vec \nabla-\vec A\right)|_n\Psi=0.
\end{equation}
This difference in boundary condition changes drastically the behavior of the
eigenvalues as function of the sample size. 
Whereas the energy of a quantum particle increases
as $E \sim 1/L^2$ for decreasing sample size $L$,
the resulting energy for the boundary condition Eq.~(3) decreases as
$E \sim L^2$. As a result 
the nucleation field decreases with increasing sample size and tends
to the limit $H_{c3}=1.695H_{c2}$ for infinitely large samples with
a flat boundary. In
circular samples this increase of $H_{c3}$ with decreasing sample size
has an oscillatory behavior due to
the quantization of the angular momentum. 
For axially symmetric samples the latter is an exact quantum number

To find the nucleation field for an 
arbitrary shaped sample we use a finite-difference representation of
the differential operator $\hat L$ on a uniform Cartesian space grid
within the link variable approach \cite{kato}. The lowest
eigenvalue and corresponding order parameter are found by applying the following
iteration procedure $\hat L\Psi^{i}=\Psi ^{i-1}$.
The calculated nucleation
field as a function of the square root of the sample area is shown in Fig.~1 for
different geometries. 
Note, that the oscillatory behavior of the
nucleation field survives even in triangular samples although the oscillations
are less pronounced then for a circular geometry.
The contour plots of the order parameter density
$|\Psi|^2$ are depicted in the insets of Fig.~1 for two square samples with different
sizes. For smaller samples areas $S=(2.32\xi)^2$,
the maximum of the order parameter is situated in the center of
the square, which corresponds to the Meissner state (see upper inset of Fig.~1).
The jump in the first derivative of the nucleation field
at $\sqrt{S}\approx 2.33\xi$ is caused by the appearance of a vortex 
in the center of the square (see lower inset of Fig.~1).
This changes the slope of the nucleation field and 
corresponds to the appearance of a vortex state with a larger value
of the angular momentum. The existence of a sequence of vortex-like states 
is inherent to all considered samples. Notice also that the nucleation field
is larger for samples with sharper corners as expected intuitively.

{\it The nucleation of superconductivity in a wedge.} 
For the wedge geometry it is more convenient to use cylindrical coordinates
($\rho$,$\phi$) with $A_{\phi}=H\rho^2/2$ and to measure all distances
in $\sqrt{c\hbar/eH}$. Then the linearized first GL equation takes the
form\cite{fomin}:
\begin{equation}
\label{eq1}
-\frac{1}{\rho}\frac{\partial}{\partial \rho}
\rho\frac{\partial \Psi}{\partial \rho}
+\frac{1}{\rho^2}\left(\frac{\partial}{\partial
\phi}-i\rho^2\right)^2\Psi=-\frac{2H_{c2}}{H} \Psi,
\end{equation}
with the boundary conditions 
\begin{equation}
\label{eq2}
\left(\frac{\partial}{\partial \phi}-i\rho^2\right)\Psi|_{\phi=0,\alpha}=0,
\quad \frac{\partial \Psi}{\partial \rho}|_{\rho\rightarrow \infty}=0.
\end{equation}
The nucleation field $H_{c3}=2H_{c2}/\lambda$ is obtained from 
the lowest eigenvalue $\lambda$
of the operator in the LHS of Eq.~(\ref{eq1}).
In order to find the lowest eigenvalue and the corresponding eigenfunction
we use again the
finite-difference technique described above. Instead of an
infinitely long wedge we consider a finite fragment with a very large radius
$R=15/\sqrt{\alpha}$ such that the value for $\lambda$
is independent of $R$.
Fig.~2 shows the spatial
distribution of the square of the absolute value of the 
order parameter $|\Psi|^2$ for two
different wedge angles $\alpha=0.05\pi$ and $0.5\pi$. For the $\alpha=0.05\pi$ case, 
the order parameter practically does not depend on the
azimuthal angle and decays faster then exponentially deep into the
sample. For $\alpha=\pi/2$, the order parameter still decays quickly
with the radius but a prominent angular dependence appears,
specially for large radii. This expected behavior, namely superconductivity
only exists in the corner \cite{fomin1},
implies a decay of the order parameter from the wedge side into the 
sample.

The numerically obtained nucleation field is shown in Fig.~3 by the
solid curve which decreases monotonically with increasing wedge
angle and diverges in the limit $\alpha\rightarrow 0$.
For $\alpha=\pi/2$, we found $H_{c3}\approx 1.96H_{c2}$ which is close
to the estimated result $H_{c3}\approx 1.82H_{c2}$ of \cite{jadallah}.
Note, that increasing the wedge angle beyond $\alpha>\pi/2$
changes very weakly
the nucleation field and we found the well-known result $H_{c3}=1.695H_{c2}$
for $\alpha=\pi$ \cite{gennes}.

Because the order parameter varies only weakly with the angle it allows us to find
the asymptotic behavior of the order parameter analytically for small wedge angles
$\alpha\ll 1$. 
To this end we rewrite Eqs.~(\ref{eq1},\ref{eq2}) as
\begin{equation}
\label{eq3}
-e^{-ix^2\eta}\frac{1}{x}\frac{\partial}{\partial x}
x\frac{\partial }{\partial x}e^{ix^2\eta}\psi
-\frac{1}{x^2\alpha^2}\frac{\partial ^2\psi}{\partial
\eta ^2}=\mu \psi,
\end{equation}
with the boundary condition
\begin{equation}
\label{eq4}
\frac{\partial \psi}{\partial \eta}|_{\eta=0,1}=0,
\end{equation}
where $x=\sqrt{\alpha}\rho$, $\eta=\phi/\alpha$, $\mu=\lambda/\alpha$,
$\psi=exp(-ix^2\eta)\Psi$ \cite{fomin}. The second term in
the LSH of Eq.~(\ref{eq3}) dominates for the case of very small wedge angles.
Therefore, the new order parameter $\psi$ depends
slightly on the angle $\eta$ which is also in agreement with the boundary
condition (\ref{eq4}). This allows to simplify the
problem. Using the boundary condition
(\ref{eq4}) we integrate Eq.~(\ref{eq3}) over
$\eta$ assuming that $\psi$ depends only on the radius  and obtain
\begin{equation}
\label{eq5}
-\frac{\partial^2\psi}{\partial x^2}
-(\frac{1}{x}+2ix)\frac{\partial\psi}{\partial x}
-(2i-\frac{4x^2}{3})\psi=\mu \psi.
\end{equation}
Note, that the wedge angle does not enter into the  last equation and the
reduced eigenvalue $\mu$ does not depend on $\alpha$.
After the substitution $\psi(x)=exp(-ix^2/2)f(x)$ Eq.~(\ref{eq5}) is
reduced to the well-known equation for the harmonic oscillator with the
lowest eigenvalue $\mu=2/\sqrt{3}$. 
Thus the order parameter for sharp wedges $\alpha
\ll 1$ can be written as
\begin{equation}
\label{eq6}
\Psi=exp\left((i\phi-i\frac{\alpha}{2}-\frac{\alpha}{2\sqrt{3}})\rho^2\right),
\end{equation}
and the nucleation field is inversly proportional to the wedge angle
\begin{equation}
\label{eq7}
H_{c3}=\frac{\sqrt{3}}{\alpha} H_{c2}.
\end{equation}
Our numerical 
results deviates from the asymptotic expression (\ref{eq7}) with about $10\%$
when the wedge angle is increased up to $\alpha \sim 0.15\pi\approx 0.5$
(compare the dashed curve with the solid curve in Fig.~3).
As is evident from Eq.~(\ref{eq3}), the corrections to the above asymptotic nucleation
field are of second order $O(\alpha^2)$.
From this observation we tried to fit $H_{c3}/H_{c2}$ to a function
$g(\alpha^2)\sqrt{3}/\alpha$. 
Within the accuracy of our numerical results we found that the nucleation
field for a wedge could be accurately fitted to
\begin{equation}
\frac{H_{c3}}{H_{c2}}=\frac{\sqrt{3}}{\alpha}
\left(1+ 0.14804\alpha^2 + \frac{0.746\alpha^2}{\alpha^2+1.8794}\right). 
\end{equation}
This function is shown by the dotted curve in Fig.~3 and agrees very well 
with our numerical results (solid curve).

Let us compare the asymptotic result for the nucleation field in
a wedge with those for thin strips and small circles. Assuming that
the order parameter depends slightly on the width $d$ of the strip and on the 
radius $R$ of the circle, we find the following asymptotic
expressions for the nucleation field in
the strip
$H_{c3}=\sqrt{3}H_{c2}\xi/d$
and the circle
$H_{c3}=2\sqrt{2}H_{c2}\xi/R$.
In conlusion, for $L\ll\xi$, where
$L$ is the smallest dimension of the sample,  
the nucleation field increases inversely with the sample size,
which corresponds to $\alpha \xi$ in the case of a wedge-like sample.

{\it Note added in proof:} The authors of Ref. \cite{fomin} recently found a
mistake in their calculation which was corrected in the erratum \cite{fominn}.
The new asymptotic behavior found with their variational calculation is 
$H_{c3}/H_{c2}=\sqrt{2/3}/\alpha=0.816/\alpha$ which is a factor
$3/\sqrt{2}=2.12$ smaller than our exact result (10).

We acknowledge discussions with J.T. Devreese and V.V. Moshchalkov. This work was supported by the
Flemish Science Foundation (FWO-Vl) and IUAP-VI. FMP is a research director
with the FWO-Vl.


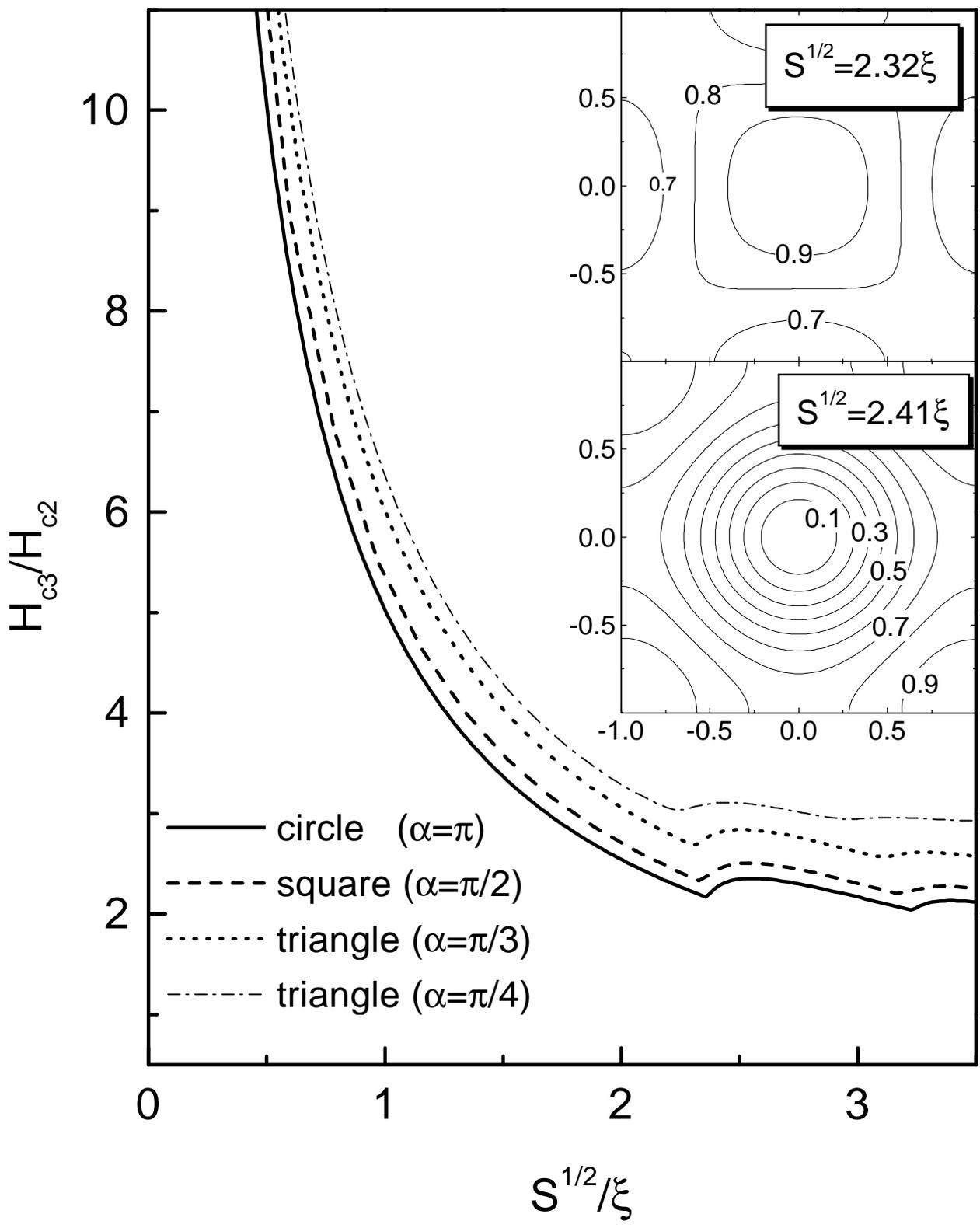
\begin{figure}
\caption{
The nucleation field as function of the square root of the sample area $S$ for
different sample geometries. The insets show $|\Psi|^2$ for  
square samples with different side lengths $\sqrt{S}$. 
}
\label{fig1}
\end{figure}

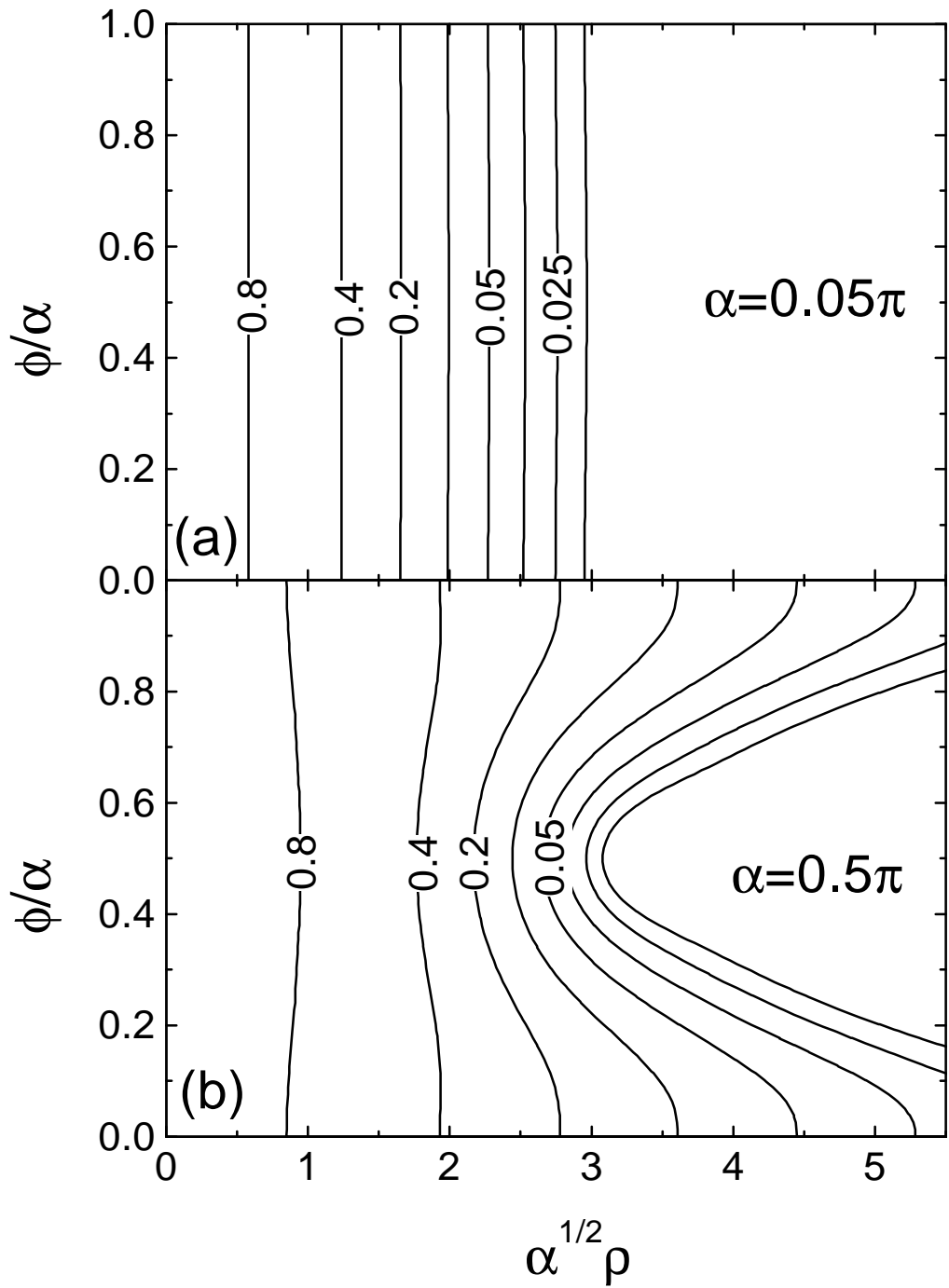
\begin{figure}
\caption{
The distribution of $|\Psi|^2$ in a wedge sample for two different
wedge angles $\alpha=0.05\pi$ (a) and $\alpha=0.5\pi$ (b).
}
\label{fig2}
\end{figure}

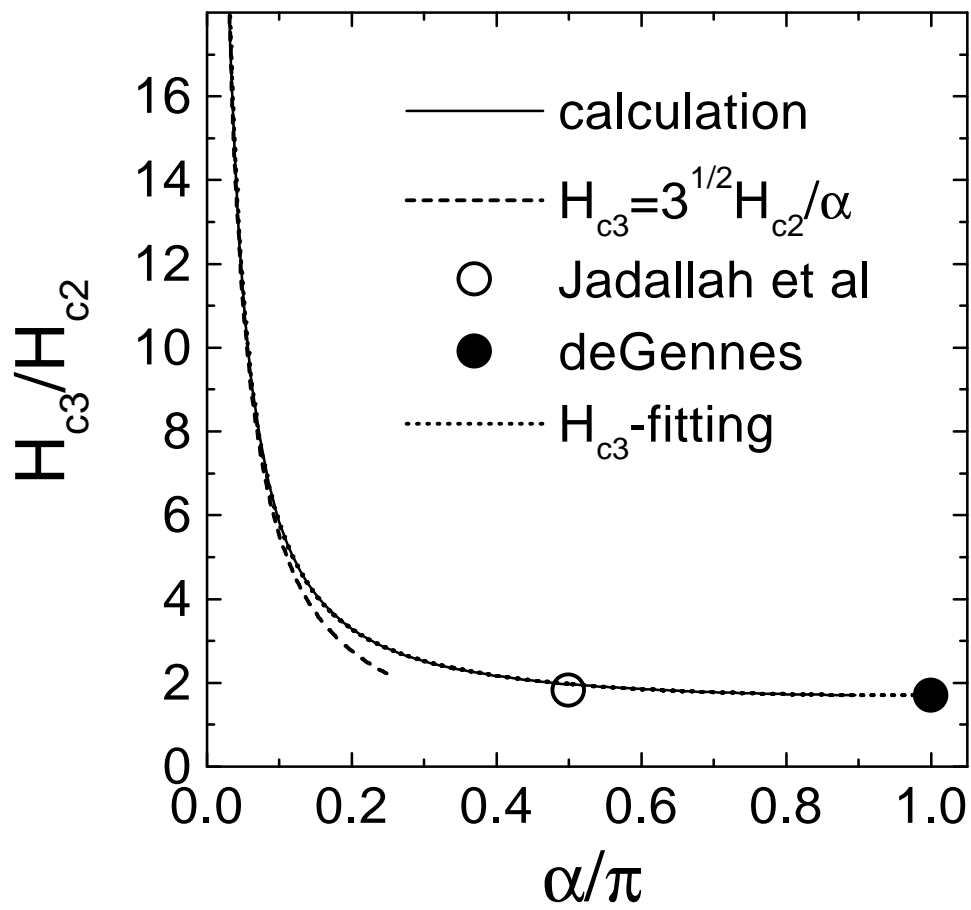
\begin{figure}
\caption{
The nucleation field in a wedge sample as function of the wedge angle
$\alpha$.
Solid curve shows the results of our finite-difference calculation,
the dashed curve corresponds to the asymptotic result for small
angles and the dotted curve is the fitted Eq.~(11). The open and solid
circles are results for $\alpha=\pi/2$ [12] and
$\alpha=\pi/2$ [16], respectively. 
}
\label{fig3}
\end{figure}

\end{document}